\documentclass[twoside]{article}
\usepackage{ijmpa1,amssymb,hyperref}
\usepackage[dvips]{graphicx}

\def\chibarchi{\left\langle\overline{\chi}\chi\right\rangle}

\begin{document}
%%%%%%%%%%%%%%%%%%%%%%%%%%%%%%%%%%%%%%%%%%%%%%%%%%%%%%%%%%%%%%%%%%%%%%%%%%%%%%%%

\runninghead{Lattice Supersymmetry with Domain Wall Fermions} {Lattice
Supersymmetry with Domain Wall Fermions}

\normalsize\textlineskip
\thispagestyle{empty}
\setcounter{page}{1}
 
\copyrightheading{}                     %{Vol. 0, No. 0 (1993) 000--000}
 
\vspace*{0.88truein}
 
\fpage{1}

\centerline{\bf LATTICE SUPERSYMMETRY}
\vspace*{0.035truein}
\centerline{\bf WITH DOMAIN WALL FERMIONS}
\vspace*{0.37truein}
\centerline{\footnotesize GEORGE T.\ FLEMING}
\vspace*{0.015truein}
\centerline{\footnotesize\it Physics Department, The Ohio State University}
\baselineskip=10pt
\centerline{\footnotesize\it Columbus, Ohio 43210-1168, USA}
%\vspace*{0.225truein}
%\publisher{(\today)}{}

\vspace*{0.21truein}
\abstracts{Supersymmetry, like Poincar\'{e} symmetry, is
softly broken at finite lattice spacing provided the gaugino mass term is
strongly suppressed.  Domain wall fermions provide the mechanism
for suppressing this term by approximately imposing chiral symmetry.
We present the first numerical simulations of ${\cal N}$=1 supersymmetric
${\rm SU}(2)$ Yang-Mills on the lattice in $d$=4 dimensions
using domain wall fermions.}{}{}

%%%%%%%%%%%%%%%%%%%%%%%%%%%%%%%%%%%%%%%%%%%%%%%%%%%%%%%%%%%%%%%%%%%%%%%%%%%%%%%%
\textlineskip
\vspace*{12pt}
%%%%%%%%%%%%%%%%%%%%%%%%%%%%%%%%%%%%%%%%%%%%%%%%%%%%%%%%%%%%%%%%%%%%%%%%%%%%%%%%

Supersymmetric (SUSY) field theories may play an important role
in describing the physics beyond the Standard Model.  Non-perturbative
numerical studies of these theories could provide confirmation
of existing analytical calculations and new insights on aspects of the theories
not currently accessible to analytic methods.  One such SUSY field theory
that can be formulated and studied numerically on the lattice is 
the four dimensional ${\cal N}$=1 super Yang-Mills (SYM), which is just QCD
with one flavor of adjoint Majorana fermions, called gluinos.
In the traditional approach, the Wilson fermions are used to simulate
the gluinos \cite{Curci:1987sm}.  Since Wilson fermions break chiral symmetry
the gluinos will be massive, breaking SUSY, unless fine-tuned mass counterterms
are introduced.  Pioneering work using these methods has already produced
very interesting numerical results \cite{Campos:1999du,Donini:1998hh}.

In this work, we summarize recent results of the first simulations
using the lattice domain wall fermion formulation to represent
the massless gluinos of SYM \cite{Fleming:2000fa,Fleming:2000bj}.  The idea
behind domain wall fermions is to start with a formulation of massless fermions
in one higher dimension and introduce domain walls so that chiral surface modes
become exponentially bound to the walls.  Pulling apart domain walls
of opposite chirality, by increasing the size of the extra dimension,
exponentially suppresses chiral symmetry breaking at a cost proportional
to the size of the extra dimension.  For reviews on DWF please see
the LATTICE '00 review talk of Vranas \cite{Vranas:2000tz}
and references therein.  The possible use of DWF in SUSY theories
has been discussed in earlier works \cite{Neuberger:1998bg,Kaplan:2000jn}
and the methods used here are along these lines.  For lists of references
not included here for lack of space, please see the cited articles
\cite{Fleming:2000fa}.

The Dirac operator in the adjoint representation of ${\rm SU}(N)$ has an index
$2N\nu$, where $\nu$ is the winding of the gauge field.  Classical instantons
have integer winding and cause condensations of operators with $2N$ gluinos
which anomalously breaks the ${\rm U}(1)$ $R$-symmetry to ${\rm Z}_{2N}$.
For $\chibarchi$ to condense, the remaining ${\rm Z}_{2N}$ symmetry must
further break either spontaneously or anomalously to ${\rm Z}_2$.
If the breaking is anomalous, then the responsible gauge configurations
must have fractional winding \cite{Fleming:2000fa}.  The existence
of such gauge configurations has already been established
\cite{Edwards:1998dj}.  It is our goal to distinguish between
these two scenarios.

All numerical simulations were performed on $8^4$ and $4^4$ Euclidean spacetime
volumes with periodic boundary conditions using the inexact hybrid molecular
dynamics (HMD) $R$ algorithm.  The algorithm numerically integrates
the classical equations of motion as part of generating a statistical ensemble
with weights proportional to the fourth root of a two adjoint flavor
Dirac determinant.  For DWF, this weight is proportional to the weight
for a single adjoint Majorana flavor.  The integration step sizes were chosen
such that systematic uncertainties due to numerical integration errors are
negligible compared to statistical uncertainties.  The $8^4$ volume
simulations were run with $\beta$$\equiv$$4/g^2$=2.3, chosen as large
as possible without entering the finite volume transition region.
The $4^4$ volume simulations were run with $\beta$=2.1.  Scaling arguments
appropriate for weak coupling suggest that the lattice spacing at $\beta$=2.1
is twice as large as at $\beta$=2.3.

To extrapolate the measured values of $\chibarchi$ to the chiral limit,
$L_s$$\to$$\infty$ and $m_f$$\to$0, simulations were performed
in $8^4$ volumes at fixed $\beta$=2.3 while the size of the extra dimension
$L_s$ was set to 12, 16, 20 or 24 and the bare mass $m_f$ was set
to 0.02, 0.04, 0.06 or 0.08.  If the formation of the gluino condensate
is due to spontaneous symmetry breaking, the lattice volume limits how small
the dynamical qluino mass can be set without losing the condensate:
$12 m_{\rm eff} \chibarchi V \gg 1$ (the 12 is just normalization)
\cite{Leutwyler:1992yt}.  As $m_{\rm eff} \gtrsim m_f$, this limit
is satisfied for all $8^4$ simulations with $m_f\ge 0.02$.

%%%%%%%%%%%%%%%%%%%%%%%%%%%%%%%%%%%%%%%%%%%%%%%%%%%%%%%%%%%%%%%%%%%%%%%%%%%%%%%%
\begin{figure}[ht]
\hfill
\begin{minipage}{0.464\textwidth}
  \includegraphics[width=\textwidth]{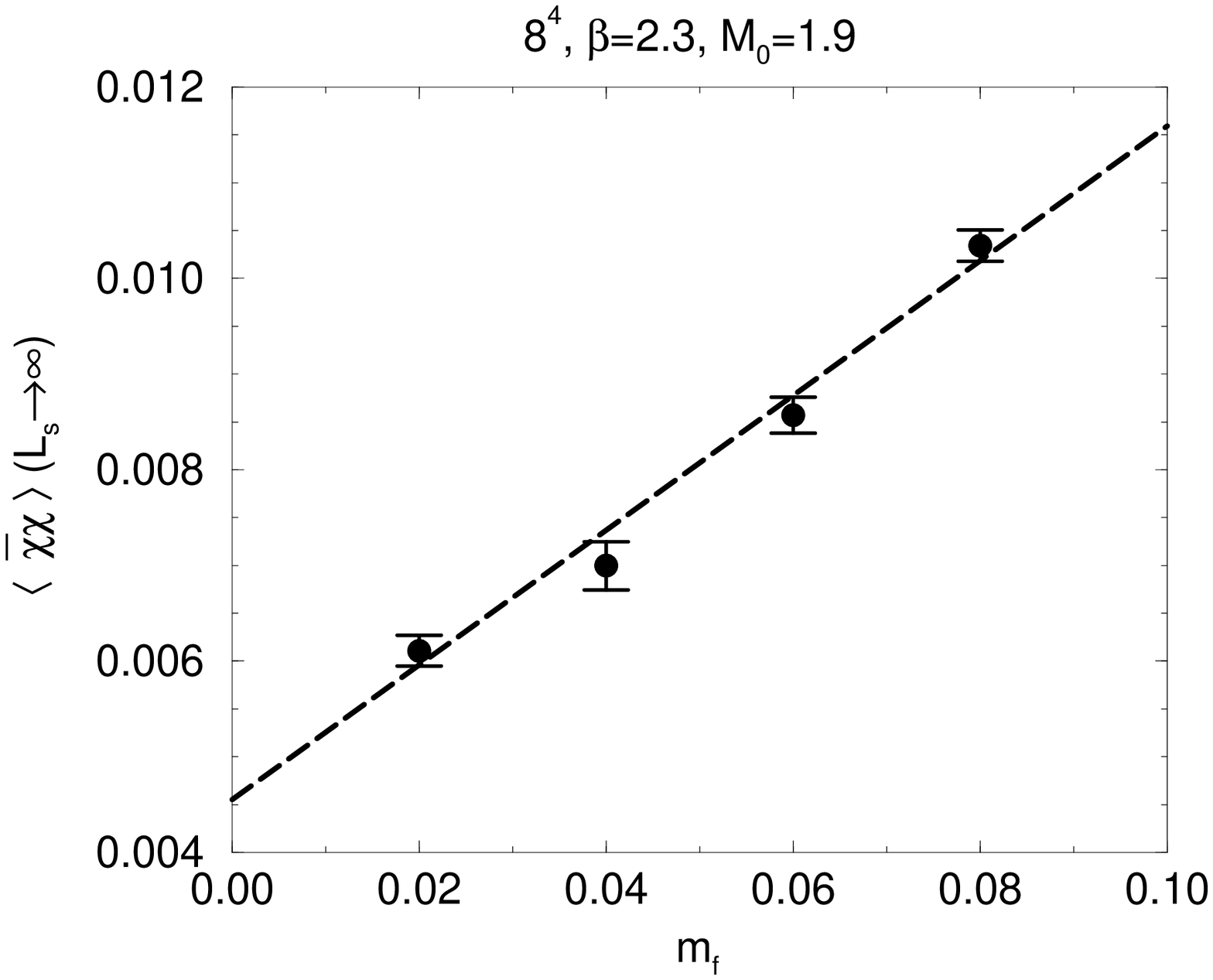}
  \vspace{-1.0cm}
  \caption{
    Extrapolated $\chibarchi$ to $L_s$$\to$$\infty$ limit {\it vs.}\ $m_f$
    and linear fit to $m_f$$\to$0 limit.
  }
  \label{fig:one}
\end{minipage}
\qquad
\begin{minipage}{0.464\textwidth}
  \includegraphics[width=\textwidth]{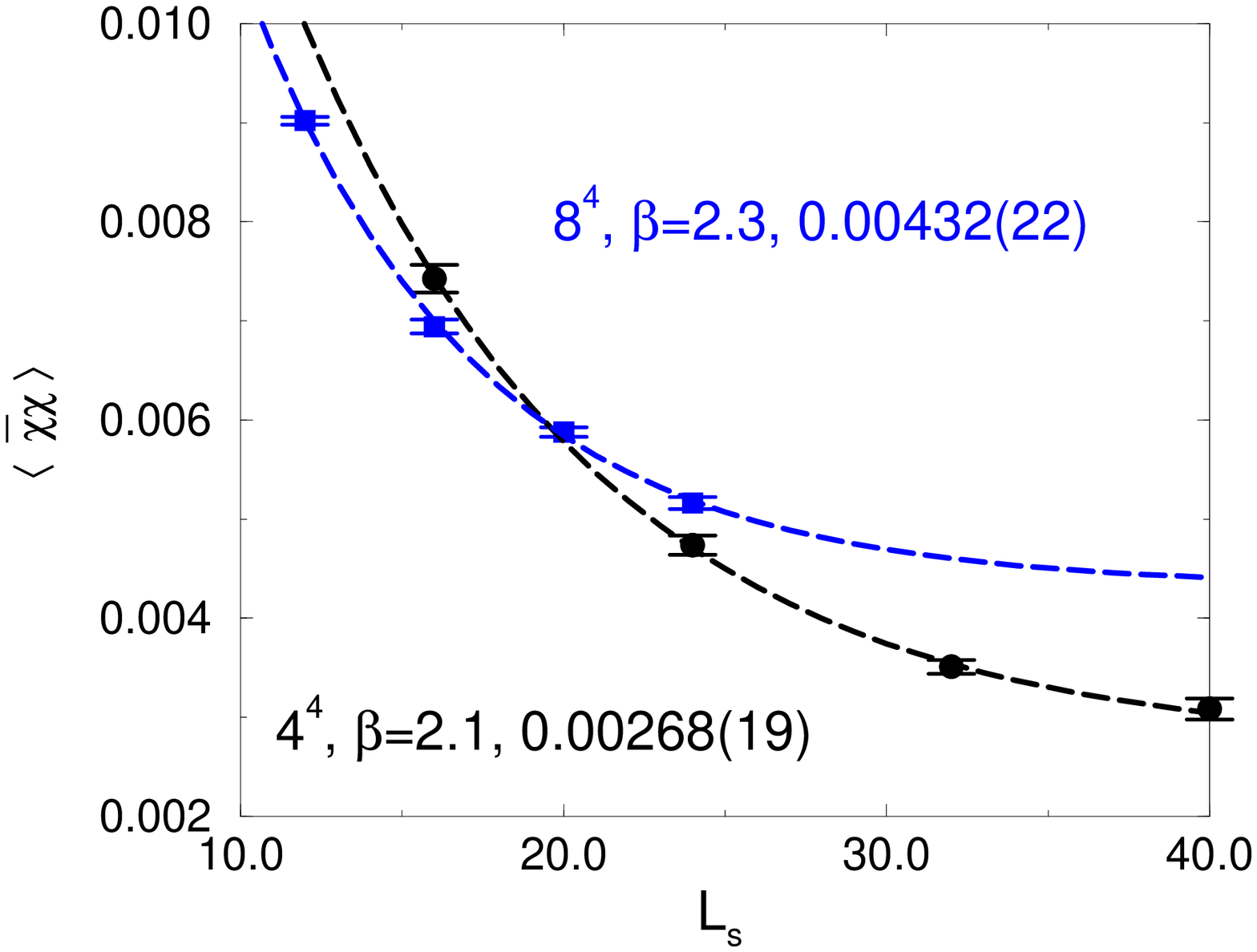}
  \vspace{-1.0cm}
  \caption{
    Dynamical $\chibarchi$ at $m_f$=0 {\it vs.}\ $L_s$ on $8^4$ and $4^4$
    lattices.  Curves are exponential fits.
  }
  \label{fig:two}
\end{minipage}
\hfill\mbox{}
\end{figure}
%%%%%%%%%%%%%%%%%%%%%%%%%%%%%%%%%%%%%%%%%%%%%%%%%%%%%%%%%%%%%%%%%%%%%%%%%%%%%%%%

To estimate the gluino condensate in the chiral limit, we first extrapolate
at fixed $m_f$ to the $L_s$$\to$$\infty$ limit using the fit function
$\chibarchi = c_0 + c_1 \exp(-c_2 L_s)$.  The values of the extrapolated
gluino condensate (with propagated errors) appear as points
in figure \ref{fig:one}.  These extrapolated values are further extraploted
to the $m_f$$\to$0 limit using a linear function
$\left.\chibarchi\right|_{L_s\!\to\!\infty} = b_0 + b_1 m_f$.  The best
fit function appears as the line in figure \ref{fig:one}, with
$b_0$=0.00455(21).  It is also reassuring to note that reversing the order
of limits, {\it i.~e.}\ first $m_f$$\to$0 at fixed $L_s$
then $L_s$$\to$$\infty$, yields a statistically consistent answer.  

Another approach to estimating the gluino condensate in the chiral limit
is to actually perform dynamical simulations with $m_f$=0.
Since finite $L_s$ will induce an exponentially small breaking of chiral
symmetry, the effective gluino mass will not be zero.  However, the gluino
mass should be too small to support spontaneous symmetry breaking.
Additional simulations were run with $L_s$ set to 12, 16, 20 or 24.  The data
are shown in figure \ref{fig:two}.  The curve is the best fit
to an exponential fitting function with the extrapolated value
of the condensate as shown.  

Surprisingly, both methods for estimating the gluino condensate produce
consistent results within the statistical errors. Note that this is
inconsistent with the notion of spontaneous symmetry breaking.  Operationally,
this result reinforces our claim that systematic uncertainties
are still relatively small despite limited statistical precision.  Further,
it gives us some confidence that our fit functions are valid over the region
of interest.

To further check for spontaneous symmetry breaking of the ${\rm Z}_4$ symmetry,
we measured $\chibarchi$ on smaller $4^4$ lattices with $m_f$=0 and
even larger values for $L_s$.  The data are shown in figure \ref{fig:two}
with the best exponential fit and the extrapolated value for the condensate.
This provides even stronger evidence that spontaneous symmetry
breaking is not responsible for the formation of a gluino condensate,
at least in finite volumes.  On these lattices $12 m_f V \chibarchi < 1$,
so analytical considerations \cite{Leutwyler:1992yt} suggest the support
of $\chibarchi$ must come primarily from topological sectors
with fractional winding of $\nu=\pm 1/2$.

The spectrum of the theory is of great interest but it was not possible
to measure it on the small lattices considered here.  Also,
the gluino condensate was measured at only two different lattice spacings
so extrapolation to the continuum limit to compare with analytical results
is not possible.  Future work could explore these very interesting topics.

%%%%%%%%%%%%%%%%%%%%%%%%%%%%%%%%%%%%%%%%%%%%%%%%%%%%%%%%%%%%%%%%%%%%%%%%%%%%%%%%

\end{document}